\documentclass[conference]{IEEEtran}

\usepackage{cite}
\usepackage{amsmath,amssymb,amsfonts}
\usepackage{algorithmic}
\usepackage{hyperref}
\usepackage{graphicx}
\usepackage{textcomp}
\usepackage{color}
\usepackage{xcolor}
\usepackage{xspace}
\usepackage{url}
\usepackage{subcaption}
\usepackage{listings}
\usepackage[frozencache,cachedir=.]{minted}
\usepackage{braket}
\usepackage{framed}
\usepackage{balance}
\usepackage{booktabs}
\usepackage[most]{tcolorbox}

\setminted{fontsize=\footnotesize}

\def\BibTeX{{\rm B\kern-.05em{\sc i\kern-.025em b}\kern-.08em
    T\kern-.1667em\lower.7ex\hbox{E}\kern-.125emX}}

\def\d4j{Defects4J\xspace}

\definecolor{codeblue}{RGB}{0, 41, 107}
\definecolor{codegrey}{RGB}{51, 51, 51}
\definecolor{codegreen}{RGB}{0, 160, 0}
\definecolor{codered}{RGB}{160, 0, 0}
\definecolor{myblue2}{RGB}{0, 43, 136}

\newcommand\inv[1]{#1\raisebox{1.15ex}{$\scriptscriptstyle-\!1$}}

\begin{document}

\title{The Inversive Relationship Between Bugs and Patches: An Empirical Study}

\author{\IEEEauthorblockN{Jinhan Kim}
\IEEEauthorblockA{\textit{School of Computing}\\
\textit{KAIST}\\
Daejeon, Republic of Korea\\
jinhankim@kaist.ac.kr}
\and
\IEEEauthorblockN{Jongchan Park}
\IEEEauthorblockA{\textit{School of Computing}\\
\textit{KAIST}\\
Daejeon, Republic of Korea\\
whdals1060@kaist.ac.kr}
\and
\IEEEauthorblockN{Shin Yoo}
\IEEEauthorblockA{\textit{School of Computing}\\
\textit{KAIST}\\
Daejeon, Republic of Korea\\
shin.yoo@kaist.ac.kr}
}

\maketitle
\thispagestyle{plain}
\pagestyle{plain}

\begin{abstract}
Software bugs\footnote{In this paper, we use `fault' and `bug'
interchangeably to refer to unwanted behaviour of a program during its
execution.} pose an ever-present concern for developers, and patching such
bugs requires a considerable amount of costs through complex operations. In
contrast, \emph{introducing} bugs can be an effortless job, in that even a
simple mutation can easily break the Program Under Test (PUT). Existing
research has considered these two opposed activities largely separately,
either trying to automatically generate realistic patches to help
developers, or to find realistic bugs to simulate and prevent future
defects. Despite the fundamental differences between them, however, we
hypothesise that they do not syntactically differ from each other when
considered simply as code changes. To examine this assumption
systematically, we investigate the relationship between
patches and buggy commits, both generated manually and automatically, using
a clustering and pattern analysis. A large scale empirical evaluation reveals
that up to 70\% of patches and faults can be clustered together based on the
similarity between their lexical patterns; further, 44\% of the code
changes can be abstracted into the identical change patterns. Moreover, we
investigate whether code mutation tools can be used as Automated Program
Repair (APR) tools, and APR tools as code mutation tools. In both cases, the
inverted use of mutation and APR tools can perform surprisingly well, or
even better, when compared to their original, intended uses. For example,
89\% of patches found by SequenceR, a deep learning based APR tool, can also
be found by its inversion, i.e., a model trained with faults and not
patches. Similarly, real fault coupling study of mutants reveals that TBar,
a template based APR tool, can generate 14\% and 3\% more fault couplings
than traditional mutation tools, PIT and Major respectively, when used as a
mutation tool. Our findings suggest that the valid scope of mining code
changes for either mutation or APR can be wider than previously thought.
\end{abstract}


\section{Introduction}
\label{sec:introduction}

Software bugs are prevalent, long-lived, and expensive to fix, thus they have 
been a major concern in software maintenance~\cite{planning2002economic,
weiss2007long}. The high cost of bug fixes mainly stems from the need for the 
developers to deeply engage with the bugs by following such steps as 1) reading 
the bug report, 2) inspecting the possible buggy locations, and 3) making 
several attempts to write patches (which involve compiling and running tests). 
All of these steps require a deep understanding of both the bug and the Program 
Under Test (PUT). In addition, developers should be aware of the possibility 
that, despite fixing the initial buggy symptoms, they may have introduced a new 
bug with their patch~\cite{yin2011fixes, le2013automatic}, further adding to 
the complexity of bug fixes.

On the contrary, we note that it is trivially easy to \emph{introduce} bugs.
Given access to the source code, bugs can be introduced effortlessly, mainly
because there exist infinitely more incorrect programs than the ones that
satisfy the given specification in the space of all programs. This fundamental
difference between writing patches and introducing bugs also can be seen in two
types of automated software testing and debugging techniques: Automated Program
Repair (APR)~\cite{goues2019automated}, which automates the process of
writing patches, and Mutation Testing~\cite{papadakis2019mutation}, which automates 
the process of introducing bugs. To successfully repair a bug, the
APR technique needs to overcome many different challenges: it first has to
accurately localise the bug, then find a specific combination of existing
ingredients to compose a patch, and finally apply and validate the patch by
executing test cases. On the contrary, mutation testing is context insensitive
and is allowed to make small syntactic changes to an arbitrary location in the
source code.

However, if we look closely at how APR tools operate, they act like mutation
tools. While trying to find a patch, it is natural for them to produce many 
\textit{faults}, which are the by-product of the search algorithms adopted by 
APR techniques. On the relationship between mutation testing and APR, Weimer et 
al. observed that Generate \& Validate (G\&V) APR techniques form a dual of mutation 
testing~\cite{Weimer2013ma}: APR aims to find mutants that pass the tests, 
while mutation testing aims to find mutants that fail the tests. One example 
that exploits this duality is PraPR~\cite{ghanbari2019practical}, a recently 
proposed APR tool that directly augments the mutation operators of a Java 
mutation testing tool, PIT~\cite{coles2016pit}. The evaluation of PraPR shows 
that the repair attempts using only PIT mutation operators perform 
surprisingly well: for \d4j v1.2.0 subjects, it produced the plausible patches for 106 
buggy versions, correct patches for 17 buggy versions out of 395 versions.


\begin{figure}[!ht]
    \centering
\begin{minted}{diff}
@@ -64,9 +64,6 @@ protected DateTimeSerializerBase(Class<T> type,
     {
         ...        
-        if (property == null) {
-            return this;
-        }
         JsonFormat.Value format = ...
         if (format == null) {
             return this;
\end{minted}
    \caption{Fix commit of JacksonDatabind-102 in \d4j}
    \label{fig:intro_null_checker_example}
\end{figure}

Given that APR and mutation testing tools have exactly opposite purposes, how
can one be successfully used as another? We hypothesise that it is mainly
because the patches and faults, when seen simply as changes made to the code,
are not that different from each other. For example, consider a code change,
from \texttt{a + b} to \texttt{a - b}. Although this is a widely used arithmetic
mutation operator, it can equally be a bug fixing patch. Similarly, even a code
change that appears to be a common fix pattern, may eventually turn out to be a
fault-inducing change, and vice versa.
Figure~\ref{fig:intro_null_checker_example} shows a such case where the
developer-written patch is a deletion of a null checker. This might seem
counter-intuitive as it is an inversion of a common fix pattern (i.e., adding a
null checker) that becomes a fault-inducing pattern, but in this case, it turns
out to be a correct patch.

Based on this observation, we design three empirical studies to investigate the
relationship between the patches and faults from different angles. First, we
compare their lexical and structural similarities by leveraging code change
clustering and pattern inferring algorithms, then evaluate whether the patches
and faults are grouped together. Second, we evaluate two mutation tools as
APR tools and evaluate their effectiveness. Finally, turning the table, we
directly convert APR tools into mutation tools and evaluate their ability to
generate mutants that are coupled with real faults. The results of our empirical
evaluation suggest that buggy and fixing code changes are in fact more similar
to each other than we expect. The implication of this finding is that, when
mining a particular type of code changes (i.e., either bug inducing commits, or
bug fixing commits), we can consider wider range of code changes than we
thought before. Further, it suggests a future direction of cross-purpose uses of
both APR and mutation testing tools, or the development of a unified technique.

The main contributions of the paper are as follows:

\begin{itemize}    
\item We empirically evaluate the similarities between patches and faults using 6k
code changes in C projects and 7k code changes in JavaScript projects,
mined from various open source repositories. The results suggest that patches
and faults are in fact similar to each other, as they can be grouped together by
the code change clustering and pattern inferring algorithms.

\item We present an empirical evaluation of mutation tools as APR techniques. We
employ mutation tools that are inversions of existing APR tools, and evaluate
these variations by applying them to the buggy programs in \d4j. The results
shows that IBIR~\cite{khanfir2020ibir} (i.e., inverted TBar~\cite{liu2019tbar})
and inverted SequenceR~\cite{chen2019sequencer} can still successfully generate 42 and 17
plausible patches, which are only eight and two fewer than their original forms.

\item We also demonstrate that existing APR tools can be converted into mutation
tools. A mutation coupling analysis using real faults in \d4j shows that TBar, a
template-based APR tool, can successfully generate 14\% and 3\% more fault
couplings than widely studied mutation tools, PIT and Major, respectively.
    
\end{itemize}

The rest of the paper is organised as follows.
Section~\ref{sec:motivating_example} introduces a motivating example.
Sections~\ref{sec:RQ1}, \ref{sec:RQ2}, and \ref{sec:RQ3} provide a detailed
experimental design and results of the three empirical studies respectively.
Section~\ref{sec:threats} discusses the threats to validity.
Section~\ref{sec:related_work} presented related previous work and
Section~\ref{sec:conclusion} concludes. 

\section{A Motivating Example}
\label{sec:motivating_example}

Figure~\ref{fig:motivating_example} contains actual code changes from \d4j.
Let us begin with a simple question: are these patches, or bug inducing changes?

\begin{figure}[!ht]
    \centering
    \begin{subfigure}{\linewidth}
        \centering
\begin{minted}{diff}
- if (dataset |\textcolor{codered}{\textbf{!=}}| null) {
+ if (dataset |\textcolor{codegreen}{\textbf{==}}| null) {
\end{minted}
        \caption{Fix change}
        \label{fig:motivating_example1}
        \vspace{1.5em}
    \end{subfigure}

    \begin{subfigure}{\linewidth}
        \centering
\begin{minted}{diff}
- if (dataset |\textcolor{codered}{\textbf{==}}| null) {
+ if (dataset |\textcolor{codegreen}{\textbf{!=}}| null) {
\end{minted}
        \caption{Inverted fix change (i.e., bug inducing change)}
        \label{fig:motivating_example2}
        \vspace{1.5em}
    \end{subfigure}

    \begin{subfigure}{\linewidth}
        \centering
\begin{minted}{diff}
|\textcolor{codegrey}{\textbf{public}}| LegendItemCollection |\textcolor{codegrey}{\textbf{getLegendItems}}|() {
    ...
    |\textcolor{codegrey}{\textbf{if}}| (|\textcolor{codegrey}{\textbf{this}}|.plot == |\textcolor{codegrey}{\textbf{null}}|) {
        return result;
    }
    |\textcolor{codegrey}{\textbf{int}}| index = |\textcolor{codegrey}{\textbf{this}}|.plot.getIndexOf(|\textcolor{codegrey}{\textbf{this}}|);
    CategoryDataset dataset = |\textcolor{codegrey}{\textbf{this}}|.plot.getDataset(index);
-   if (dataset |\textcolor{codered}{\textbf{!=}}| null) {
+   if (dataset |\textcolor{codegreen}{\textbf{==}}| null) {
        |\textcolor{codegrey}{\textbf{return}}| result;
    }
    |\textcolor{codegrey}{\textbf{int}}| seriesCount = dataset.getRowCount();
    ...
\end{minted}
        \caption{Fix change with surrounding context}
        \label{fig:motivating_example3}        
    \end{subfigure}

    \caption{A fix commit of jfreechart (Chart-1 in \d4j)}
    \label{fig:motivating_example}
\end{figure}

Figure~\ref{fig:motivating_example1} contains an example fix change from
jfreechart that modifies \texttt{!=} to \texttt{==}, while
Figure~\ref{fig:motivating_example2} shows the inversion of
Figure~\ref{fig:motivating_example1}. Without the captions, it is not easy to
tell them apart. We hypothesise that a code change itself is context
insensitive, which is why it is hard to distinguish a patch from a bug inducing
change, and vice versa. The difference becomes clearer only when we consider the
surrounding context of the change, as shown in
Figure~\ref{fig:motivating_example3}. An earlier conditional statement
\mintinline[escapeinside=||]{diff}{|\textcolor{codegrey}{\textbf{if}}|
(|\textcolor{codegrey}{\textbf{this}}|.plot ==
|\textcolor{codegrey}{\textbf{null}}|)} has a similar predicate and the same
return statement; we also observe that the variable \texttt{dataset}
is subsequently used in Line 13, suggesting that it is more natural to return
when it is
\texttt{null}. With
the help from the context, we can guess that the change in
Figure~\ref{fig:motivating_example1} is likely to be a fixing change, and not a
bug inducing one. This example raises the questions of whether mining and
interpreting the given code changes only as either fix or bug inducing is
desirable, as well as how actually different they are from each other. In the
following three sections, we investigate and discuss those observations with three
empirical studies, respectively.

\begin{figure}[!ht]
    \centering
    \begin{subfigure}{\linewidth}
        \centering
\begin{minted}{diff}
@@
|\textcolor{codegrey}{\textbf{expression}}| E0;
@@
- if (isupper(|\textcolor{codered}{\textbf{(int )}}|*E0))
+ if (isupper(*E0))
  {
-   *E0 = tolower(|\textcolor{codered}{\textbf{(int )}}|E0);
+   *E0 = tolower(*E0);
  }
\end{minted}
        \caption{IfStatement/21/1/0}
        \label{fig:spinfer_example1}
    \end{subfigure}

    \begin{subfigure}{\linewidth}
        \centering
\begin{minted}{diff}
@@
|\textcolor{codegrey}{\textbf{expression}}| E0;
@@
- if (isdigit(|\textcolor{codered}{\textbf{(int )}}|*E0))
+ if (isdigit(*E0))
  {
   ...
  }
\end{minted}
        \caption{IfStatement/21/1/1}
        \label{fig:spinfer_example2}
    \end{subfigure}
    
    \caption{Example clusters resulted from FlexiRepair}
    \label{fig:spinfer_example}
\end{figure}

\section{Similarity Study (RQ1)}
\label{sec:RQ1}

\textbf{RQ1. How similar are patches and faults to each other?}
We answer RQ1 using two approaches: clustering both patches and faults, and
abstracting both as change patterns. With clustering, we investigate whether
patches and faults can belong to the same cluster, whereas with pattern
inferring, we investigate whether patches and faults can be abstracted into the
same pattern. Answers to these questions would provide evidence of how much the
patches and faults have similar structures and patterns. We perform the
clustering analysis using FlexiRepair~\cite{koyuncu2020flexirepair}, and the
pattern inferring analysis using SemSeed~\cite{patra2021semantic}.






\subsection{Cluster Analysis with FlexiRepair}
\label{sec:RQ1_flexirepair}

FlexiRepair~\cite{koyuncu2020flexirepair} is an extension and combination of
FixMiner~\cite{koyuncu2020fixminer} and Spinfer~\cite{serrano2020spinfer}.
Below, we will briefly introduce them with the procedure of how the clusters are
formed.

\subsubsection{FixMiner}
To represent a code change, FixMiner~\cite{koyuncu2020fixminer} constructs a
tree representation with three types of information: \texttt{Shape},
\texttt{Action}, and \texttt{Token}. Each of the information type corresponds to
an abstraction level of the resulting representation. FixMiner starts the
clustering at the highest abstraction level, which is \texttt{Shape}, and
successively applies clustering to the results from the previous abstract level
with \texttt{Action} and \texttt{Token}.\footnote{As FlexiRepair only considered
\texttt{Shape} and \texttt{Action} for clustering, we exclude \texttt{Token}
from the clustering process.} First, at abstraction level of \texttt{Shape},
FixMiner groups the code changes based on the type of the root node of AST and
its depth (e.g., IfStatement/7), and identifies their clusters using
algorithms of GumTree~\cite{falleri2014fine}. Next, the abstraction level
of \texttt{Action} further divides the clusters generated by \texttt{Shape},
based on the change operations identified by GumTree. These clusters are
labelled as node/depth/ShapeTreeClusterId (e.g., IfStatement/7/2).

\subsubsection{Spinfer}
Spinfer~\cite{serrano2020spinfer} aims to infer semantic patches of Linux kernel
by identifying the similar code fragments and control flows across code changes.
Following FlexiRepair~\cite{koyuncu2020flexirepair}, we use Spinfer to finalise
the clustering process on each cluster by FixMiner. As a result, the clusters
generated by Spinfer have the lowest abstraction level and indexed as
node/depth/ShapeTreeClusterId/ActionTreeClusterId (e.g., IfStatement/7/2/0).

The final output of Spinfer is the clusters of generic patches in the form of 
Coccinelle~\cite{lawall2018coccinelle} transformation rules: 
Figure~\ref{fig:spinfer_example} shows two of generated clusters. They are 
assigned to the same cluster of IfStatement/21/1 at the \texttt{Action} 
abstraction level by FixMiner, but have been further divided into two smaller 
clusters, IFStatement/21/1/0 and IFStatement/21/1/1, by Spinfer because they 
have different structures of body of if statement.

\begin{table}[!ht]
    \centering
    \caption{C subject programs for cluster analysis using FlexiRepair}
    \label{tab:RQ1_dataset}
    \begin{tabular}{lr}
        \toprule
        Repository        & \# Commits  \\
        \midrule
        libtiff           & 3,570      \\ 
        cmake             & 38,414     \\ 
        redis             & 8,770      \\ 
        gzip              & 604        \\ 
        libarchive        & 5,472      \\ 
        cairo             & 11,724     \\ 
        curl              & 26,967     \\ 
        tcl               & 17,145     \\ 
        nginx             & 6,858      \\ 
        apr               & 8,945      \\ 
        openssh-portable  & 11,004     \\
        gmp               & 16,782     \\
        lighttpd1.4       & 3,882      \\
        lighttpd2         & 1,551      \\
        git               & 46,715     \\
        MonetDBLite-C     & 48,461     \\
        freeradius-server & 37,535     \\
        bind9             & 31,286     \\
        tmux              & 7,790      \\
        gstreamer         & 19,016     \\
        
        \bottomrule
    \end{tabular}
\end{table}

\subsubsection{Dataset}
We reuse the dataset presented by FlexiRepair: it contains mined commits in C
projects from  Github, Gitlab, and Savannah. However, it was not possible for us
to process all repositories provided by the FlexiRepair dataset due to the
limited computational resources. Instead, we randomly selected 20 repositories,
as listed in Table~\ref{tab:RQ1_dataset}, and collect the code changes with the
inverted changes, resulting in 720,000 changes. After applying the filtering
rules of FlexiRepair to extract only fix commits, we have 3,000 fix changes that
result in total 6,000 code changes considering inverted changes (i.e., bug
inducing changes).

\subsubsection{Evaluation} 
For the evaluation of FlexiRepair clusters, we use three levels of clusters from
\texttt{Shape}, \texttt{Action}, and Spinfer, respectively. For the sake of simplicity, we
hereafter denote the three levels as `Level1', `Level2', and `Level3'. We report
the number of clusters that include both the patches and faults for each level.
Note that we cannot report other evaluation metrics such as cluster membership
accuracy, as there is no ground truth of correct cluster membership for code
changes.

\subsection{Pattern Analysis with SemSeed}
\label{sec:RQ1_semseed}

SemSeed~\cite{patra2021semantic} presents an efficient algorithm for seeding
realistic bugs by abstracting and matching bug inducing patterns.
SemSeed selects an AST subtree of the
changed line and extracts two token sequences (before and after the change) from
the subtree. From them, the bug seeding patterns are inferred by abstracting all
identifiers and literals into the placeholders, resulting in a pair of
abstracted token sequences, $t = \braket{t_{x}, t_{y}}$. The tokens in $t_{x}$
and $t_{y}$ are either identifier or literals, or non-identifiers and
non-literals.

As SemSeed only learns its patterns from faults, we modify it to consider both
patches and faults, and see how many same $\braket{t_{x}, t_{y}}$ pairs are
found in both of them. Let $t \in T_{p}$ be the pairs from the patches and $t
\in T_{f}$ be the pairs from the faults. We report the number of pairs that
exist in both $T_{p}$ and $T_{f}$, which would be the code changes that
represent both patches and faults. Based on the dataset presented by SemSeed
that contains 3,600 inverted fix changes from 100 JavaScript projects in
Github, we build our own that includes both fix and inverted changes, resulting in total 7,200 code
changes. 

\begin{table}[!ht]
    \centering
    \caption{Cluster analysis with FlexiRepair. $x$ refers to the number
        of clusters having both patches and faults and $y$ refers to the
        number of total clusters.}
    \label{tab:RQ1}
    \begin{tabular}{l|lll}
        \toprule
        Cluster  & Level1           & Level2               & Level3             \\
             level       & (node/depth)     & (ShapeTreeClusterId) & (SpinferClusterId) \\
        \midrule
        $x$ / $y$     & 222 / 316 (70\%) & 569 / 1,800 (31\%)   & 227 / 3,859 (7\%)  \\
        \bottomrule
    \end{tabular}
\end{table}

\begin{figure}[!ht]
    \begin{subfigure}{\linewidth}
        \centering
\begin{minted}{diff}
@@
|\textcolor{codegrey}{\textbf{identifier}}| I0;
|\textcolor{codegrey}{\textbf{expression}}| E1;
@@
- char *I0 = rad_malloc(E1 |\textcolor{codered}{\textbf{+ 1}}|);
+ char *I0 = rad_malloc(E1);
\end{minted}
        \caption{DeclStmt/10/0/0}
        \label{fig:RQ1_example_level1_a}
    \end{subfigure}
    \begin{subfigure}{\linewidth}
        \centering
\begin{minted}{diff}
@@
|\textcolor{codegrey}{\textbf{identifier}}| I0;
|\textcolor{codegrey}{\textbf{expression}}| E1;
@@
- |\textcolor{codered}{\textbf{const}}| char *I0 = E1;
+ char *I0 = |\textcolor{codegreen}{\textbf{(char *)}}|E1;
\end{minted}
        \caption{DeclStmt/10/18/2}
        \label{fig:RQ1_example_level1_b}
    \end{subfigure}

    \caption{Example Level1 cluster (DeclStmt/10)}
    \label{fig:RQ1_example_level1}
\end{figure}

\begin{figure}[!ht]
    \begin{subfigure}{\linewidth}
        \centering
\begin{minted}{diff}
@@
|\textcolor{codegrey}{\textbf{identifier}}| I1 = {getch,strequal};
|\textcolor{codegrey}{\textbf{expression list}}| E2;
|\textcolor{codegrey}{\textbf{expression}}| E0;
@@
- E0 = I1(E2); 
+ E0 = |\textcolor{codegreen}{\textbf{(char )}}|I1(E2); 
\end{minted}
        \caption{ExprStmt/5/0/1}
        \label{fig:RQ1_example_level2_a}
    \end{subfigure}
    \begin{subfigure}{\linewidth}
        \centering
\begin{minted}{diff}
@@
|\textcolor{codegrey}{\textbf{assignment operator}}| A1;
|\textcolor{codegrey}{\textbf{expression}}| E2, E0, E3;
@@
- E0 A1 strlen(*E2) + E3; 
+ E0 A1 |\textcolor{codegreen}{\textbf{(int )}}|strlen(*E2) + E3; 
\end{minted}
        \caption{ExprStmt/5/0/2}
        \label{fig:RQ1_example_level2_b}
    \end{subfigure}

    \caption{Example Level2 cluster (ExprStmt/5/0)}
    \label{fig:RQ1_example_level2}
\end{figure}

\subsection{Results of Cluster Analysis with FlexiRepair}
\label{sec:RQ1_results_1}

Table~\ref{tab:RQ1} shows the number of clusters that have both patches and 
faults ($x$) and the total number of clusters ($y$). Among all clusters, 70\% 
of Level1 clusters contain both patches and faults, followed by 31\% of Level2 
clusters, and 7\% of the Level3 clusters: the trend confirms our expectation 
that the less we abstract the code changes, the more separated they would be.
Figures~\ref{fig:RQ1_example_level1} and \ref{fig:RQ1_example_level2}
present a closer look at how clusters are formed at Level1 and Level2:
Figure~\ref{fig:RQ1_example_level1_a} and Figure~\ref{fig:RQ1_example_level1_b} 
are in same cluster of DeclStmt/10, and Figure~\ref{fig:RQ1_example_level2_a} 
and Figure~\ref{fig:RQ1_example_level2_b} are in same cluster of ExprStmt/5/0.
Both levels have placeholder identifiers or expressions that can be replaced,
but Level2 clusters tend to show closer resemblance.

\begin{figure}[!t]
    \includegraphics[width=\linewidth]{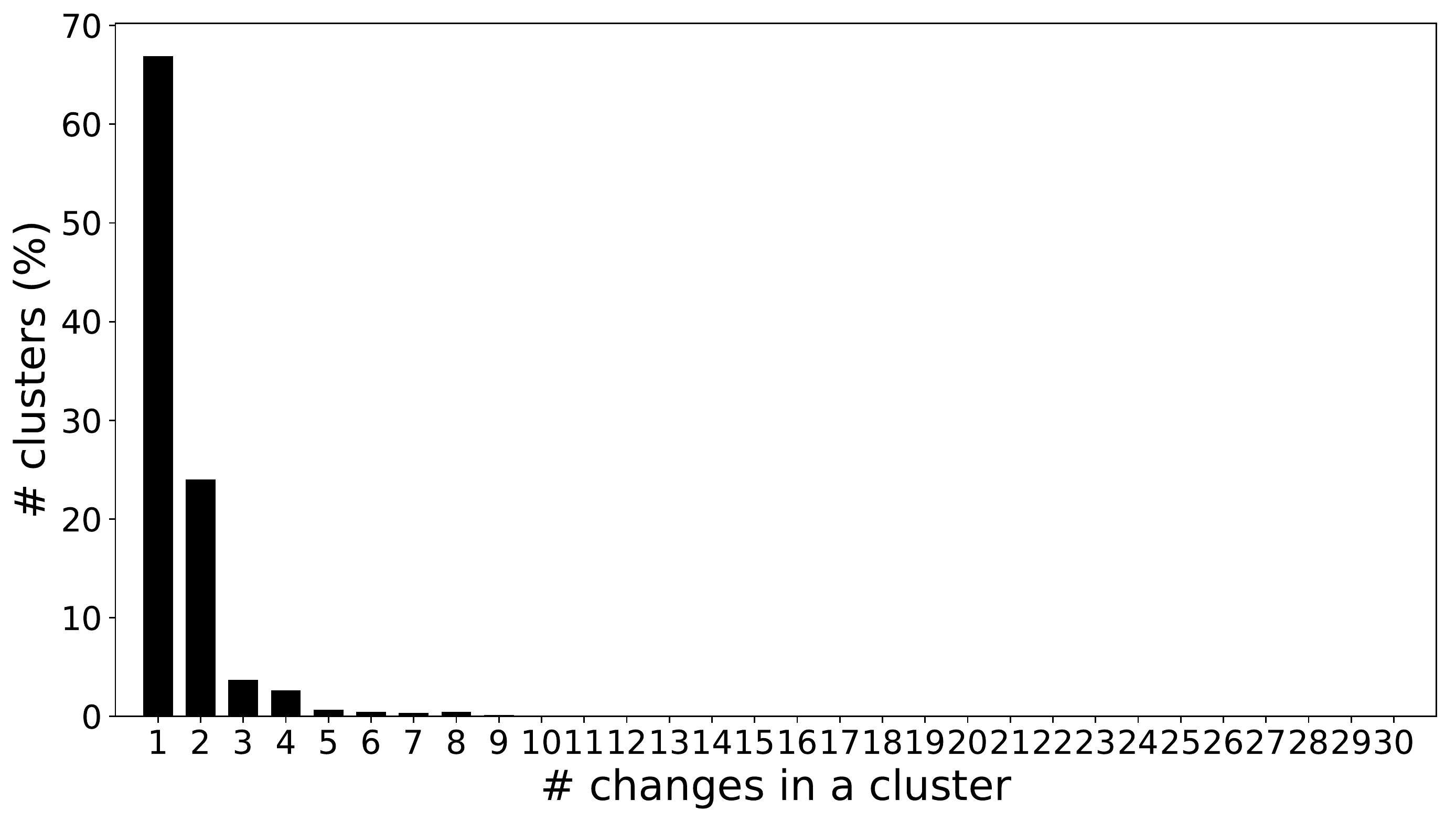}
    \caption{The number of code changes for each Level3 cluster.}
    \label{fig:RQ1_cluster_frequency}
\end{figure}

\begin{figure}[!ht]
    \begin{subfigure}{\linewidth}
        \centering
\begin{minted}{diff}
- reject(error);
+ reject(|\textcolor{codegreen}{\textbf{convertImapError}}|(error));
\end{minted}
        \caption{Fix commit of Adobe Brackets @a460c47}
        \label{fig:RQ1_semseed_example_a}
        \vspace{1em}
    \end{subfigure}
    \begin{subfigure}{\linewidth}
        \centering
\begin{minted}{diff}
- callback(_err);
+ callback(|\textcolor{codegreen}{\textbf{mapError}}|(_err));
\end{minted}
        \caption{Inverted fix commit of nylas-mail @93942e7}
        \label{fig:RQ1_semseed_example_b}
    \end{subfigure}

    \caption{Two code changes having same SemSeed patterns}
    \label{fig:RQ1_semseed_example}
\end{figure}

In addition, we examine how many code changes are in each Level3 cluster,
since the overlaps of the patches and faults in Level3 clusters are only 227 
out of 3,859 cases (7\%). Figure~\ref{fig:RQ1_cluster_frequency} shows that 67\%
of Level3 clusters contain only one code change, failing to be grouped with any 
others due to the low abstraction level. Once we exclude such singletons, the 
proportion of the clusters that have both the patches and faults becomes 
22\%.


\subsection{Results of Pattern Analysis with SemSeed}
\label{sec:RQ1_results_2}

Next, we analyse the patterns of the patches and faults using SemSeed. We count
the code changes whose patterns ($t$) are found in both patches ($T_{p}$) and
faults ($T_{f}$).
Out of 3,951 code changes for each group, 1,752 code changes (44\%) have
patterns found in both $T_{p}$ and $T_{f}$. For example,
Figure~\ref{fig:RQ1_semseed_example} shows two code changes from different
projects: Figure~\ref{fig:RQ1_semseed_example_a} is a fix change from Adobe
Brackets which adds a method call \texttt{convertImapError} around
\texttt{error}, and Figure~\ref{fig:RQ1_semseed_example_b} is a bug inducing
change from nylas-mail which does similar modifications to
Figure~\ref{fig:RQ1_semseed_example_a}. Both changes are abstracted into the
pattern from $Idf_1(Idf_3);$ to $Idf_1 (Idf_2 (Idf_3));$ where $Idf$ represents
a placeholder identifier.

\begin{tcolorbox}[boxrule=0pt,frame hidden,sharp corners,enhanced,borderline north={1pt}{0pt}{black},borderline south={1pt}{0pt}{black},boxsep=2pt,left=2pt,right=2pt,top=2.5pt,bottom=2pt]
\textbf{Answer to RQ1:} Up to 70\% of patches and faults can be clustered 
together by FlexiRepair; 44\% of them can be abstracted into the same pattern 
by SemSeed. We conclude that patches and faults are not mutually exclusive and 
can be similar to each other.
\end{tcolorbox}


\begin{figure}[!ht]
    \begin{subfigure}{0.5\linewidth}
        \centering
        \includegraphics[width=\linewidth]{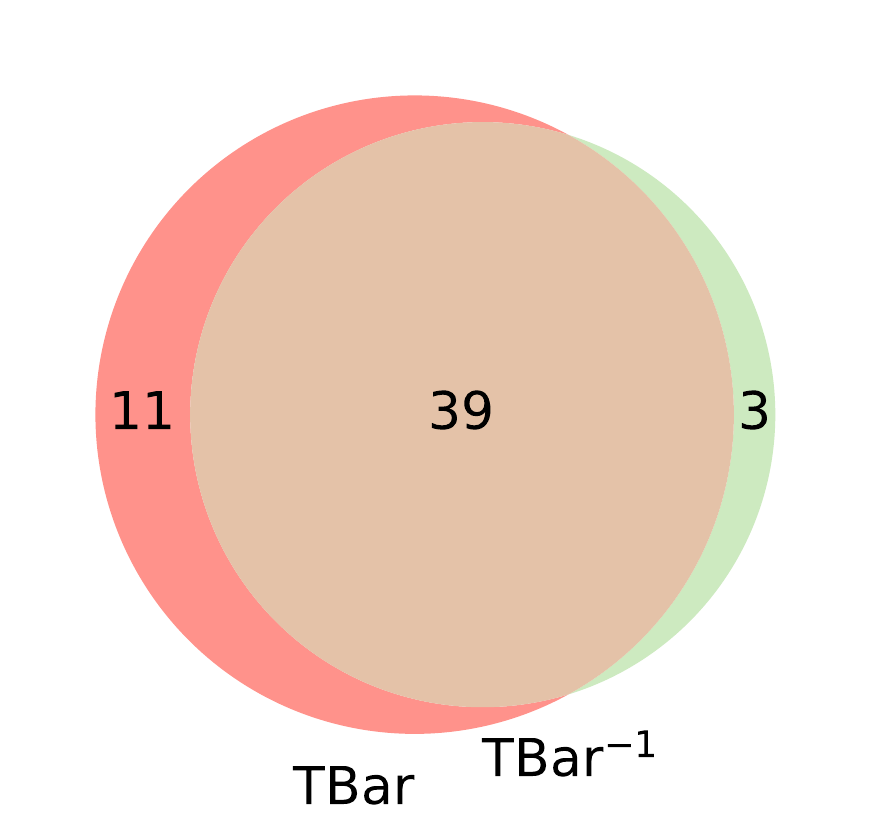}
        \caption{TBar and \inv{TBar}}
        \label{fig:RQ2_venn_TBar}
    \end{subfigure}%
    \begin{subfigure}{0.5\linewidth}
        \centering
        \includegraphics[width=\linewidth]{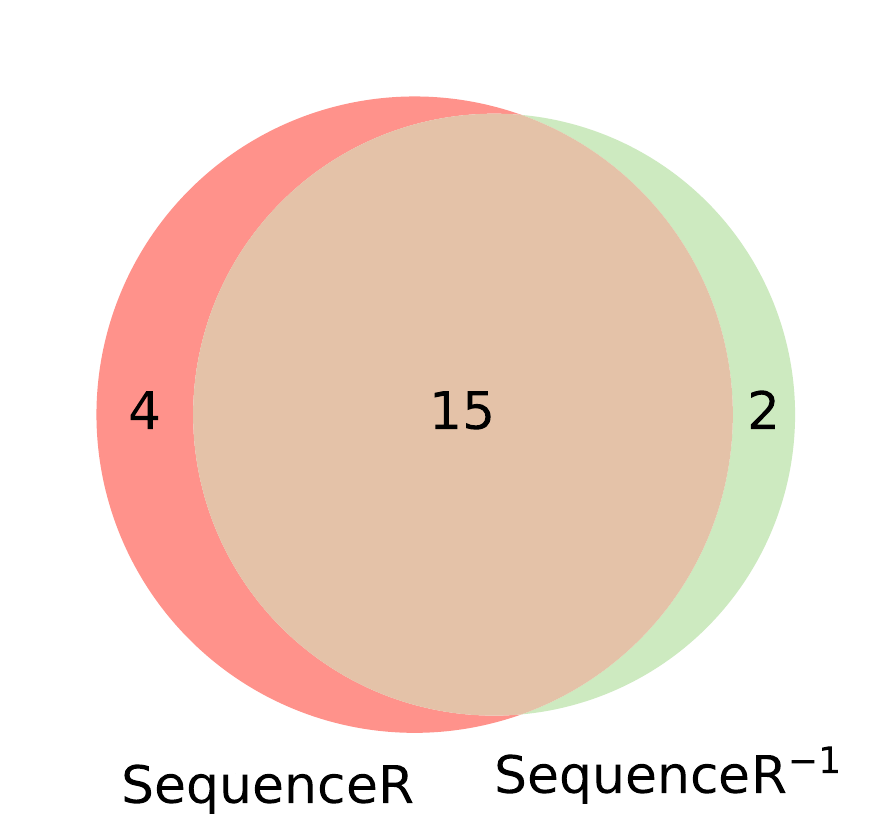}
        \caption{SequenceR and \inv{SequenceR}}
        \label{fig:RQ2_venn_SequenceR}
    \end{subfigure}
\caption{The number of bugs for which mutation or APR can generate plausible patches.}
\label{fig:RQ2_venn}
\end{figure}

\section{Mutation-to-APR Study (RQ2)}
\label{sec:RQ2}

\textbf{RQ2. Can mutation tools be APR tools?} 
Given that the patches and faults have similar forms, we now move on to the
cross-evaluation between them by investigating the patch effectiveness of the
mutation. We treat the generated faults as patches and see whether they are
actually plausible patches. If the patches and faults are fundamentally
different and mutually exclusive, mutation tools will fail to generate any
plausible patches.

Ideally, we would use existing mutation tools to perform this study.
However, among the available and widely used Java mutation tools,
PIT~\cite{coles2016pit} has already been evaluated as an APR tool by
PraPR~\cite{ghanbari2019practical}, whereas Major~\cite{just2014major} does not
allow any modifications that are necessary for our experimentation as its source
code is not available. As a result, we use IBIR~\cite{khanfir2020ibir}, which
mutates source code using inverted templates of TBar~\cite{liu2019tbar}, and
\inv{SequenceR}, which is originally the Neural Machine Translation (NMT) based
APR tool, SequenceR~\cite{chen2019sequencer}, but trained with inverted dataset
so that it `translates' correct code to buggy code. Essentially, these two tools
are direct inversions of APR tools, meaning that they are now designed to
introduce bugs. Consequently, if these inversions can still patch bugs, it would
show that our hypothesis about the similarities between patches and bugs is
correct.

\subsection{Modification Details}
\label{sec:RQ2_modification}

For this study, we introduce the following modifications to the original tools.
\begin{itemize}
\item \textit{IBIR:} To use IBIR as an APR tool, we take the source code of 
TBar and replace the fix templates of TBar with the templates of IBIR. We 
switch templates instead of directly using IBIR, in order to take advantage of 
APR related support functionalities that already reside in TBar. We denote this 
modified tool \inv{TBar} to differentiate it from the original TBar as well 
as IBIR.

\item \textit{SequenceR:} SequenceR learns to translate a buggy code into a 
correct code using the fix changes. We follow the same training procedure, but 
we place the buggy code in the position of the correct code, and vice versa: 
this model will learn how to translate the correct code to the buggy code. We 
denote this model \inv{SequenceR}. 
\end{itemize}

\subsection{Evaluation Metrics and Configurations}
\label{sec:RQ2_metric}

We report and compare the number of bugs for which mutation and APR tools
generate plausible patches. We use \d4j v1.2.0, which includes four project and
231 bugs.\footnote{See details at
\url{https://github.com/rjust/defects4j/tree/v1.2.0}.} We use the old version of
\d4j, v1.2.0, as some of the studied tools are specifically designed to target
\d4j v1.2.0. Note that we compare SequenceR and \inv{SequenceR} using only 75
bugs, as the original tool is designed to target only one line patches.

\subsection{Results}
\label{sec:RQ2_results}

We run two mutation tools, \inv{TBar} and \inv{SequenceR}, and two corresponding
APR tools, TBar and SequenceR, on the buggy programs in \d4j.
Figure~\ref{fig:RQ2_venn} depicts two Venn diagrams that show the number of bugs
for which each tool can generate plausible patches. TBar generates plausible
patches for 50 bugs, which is eight more than those fixed by \inv{TBar};
SequenceR generates plausible patches for 19 bugs, which is two more than those
fixed by \inv{SequenceR}. There are overlaps of 73\% (39) between TBar and
\inv{TBar}, and 71\% (15) between SequenceR and \inv{SequenceR}, respectively.
Although, APR tools are better than mutation tools at generating plausible
patches in general, the results suggest that mutation tools can successfully
perform as APR tools even though they learnt to operate in the opposite
direction. 

We also perform a qualitative analysis of the bugs that
either APR or mutation tool exclusively fixes. Of 11 bugs fixed only by TBar,
eight are fixed with the insertion operators, five of which are related to adding
null pointer checkers. On the other hand, out of three bugs fixed only by
\inv{TBar}, two are fixed with the deletions operators. We suspect
that this is due to the difference in the number of operators in these tools:
TBar includes 12 insertion operators but only two deletion operators, meaning
that \inv{TBar} has 12 deletion operators and two insertion operators.

SequenceR exclusively fixes four bugs, three of which are fixed with the
variable replacements. The two exclusive fixes by \inv{SequenceR} are also related
to the variable replacement and changing the condition of a if statement. Since
SequenceR is an NMT based tool, it is difficult to analyse and interpret its
operations. Further analysis would require improved explainability of the
underlying NMT models.






\begin{tcolorbox}[boxrule=0pt,frame hidden,sharp corners,enhanced,borderline north={1pt}{0pt}{black},borderline south={1pt}{0pt}{black},boxsep=2pt,left=2pt,right=2pt,top=2.5pt,bottom=2pt]
\textbf{Answer to RQ2:} Despite having learnt from faults, mutation tools can
successfully find plausible patches for 42 and 17 buggy programs in \d4j, 
compared to their counterpart APR tools that find plausible patches for 50 and 19
buggy programs.

\end{tcolorbox}

\section{APR-to-Mutation Study (RQ3)}
\label{sec:RQ3}

\textbf{RQ3. Can APR tools be mutation tools?} We investigate whether the code
changes produced by APR tools can be used as bug injections, despite the
original intention of being patches. Conducting a real fault coupling study
by following Just et al.~\cite{just2014mutants}, we evaluate the effectiveness of
mutants produced by APR tools. As there are inherent differences between APR
tools and mutation tools, we describe the challenges we faced, as well as how we
dealt with them.



\subsection{What kinds of APR tools can we modify?}

Table~\ref{tab:apr_tools} presents a list of Java APR tools we have considered.
Our final selection criteria are as follows:

\begin{itemize}
\item \textit{Availability}: we exclude the tools that are not publicly
available or do not make their source code publicly available.
Hercules~\cite{saha2019harnessing}, CapGen~\cite{wen2018context}, and
CURE~\cite{jiang2021cure} are excluded for this reason.

\item \textit{Executability}: We exclude ssFix~\cite{xin2017leveraging} because
it fails to connect to the private code search engine, and
CoCoNut~\cite{lutellier2020coconut} because it has unresolved issues in
preprocessing of training data as well as the model
training.\footnote{\url{https://github.com/lin-tan/CoCoNut-Artifact/issues/2}}

\item\textit{Failing tests}: We exclude the tools that require failing tests
for patch generation, because we will apply them to correct programs for
injecting faults. We exclude APR tools based on genetic programming, such as
GenProg~\cite{le2011genprog} and ARJA~\cite{yuan2018arja},
because they use fitness functions that check whether tests that originally
failed subsequently pass.
\end{itemize}

After filtering, we are left with four APR tools: SimFix~\cite{jiang2018shaping},
PraPR~\cite{ghanbari2019practical}, TBar~\cite{liu2019tbar}, and
SequenceR~\cite{chen2019sequencer}. These have all been recently published and
open sourced; further, they do not require failing tests.

\begin{table}[!h]
    \centering
    \caption{Considered Java APR tools}
    \label{tab:apr_tools}
    \begin{tabular}{lrrrr}
        \toprule
        Tool                                & Selected? & Public? & Working? & Failing  \\
                                            &           &         &          & tests      \\
        \midrule

        GenProg~\cite{le2011genprog}        & No        & Yes     & Yes      & Yes       \\
        Angelix~\cite{mechtaev2016angelix}  & No        & Yes     & Yes      & Yes       \\
        Nopol~\cite{xuan:hal-01285008}      & No        & Yes     & Yes      & Yes       \\
        ssFix~\cite{xin2017leveraging}      & No        & Yes     & No       & No        \\
        CapGen~\cite{wen2018context}        & No        & No      & -        & No        \\
        ARJA~\cite{yuan2018arja}            & No        & Yes     & Yes      & Yes       \\
        SketchFix~\cite{hua2018towards}     & No        & Yes     & Yes      & Yes       \\
        SimFix~\cite{jiang2018shaping}      & Yes       & Yes     & Yes      & No        \\
        Hercules~\cite{saha2019harnessing}  & No        & No      & -        & No        \\
        PraPR~\cite{ghanbari2019practical}  & Yes       & Yes     & Yes      & No        \\
        TBar~\cite{liu2019tbar}             & Yes       & Yes     & Yes      & No        \\
        SequenceR~\cite{chen2019sequencer}  & Yes       & Yes     & Yes      & No        \\
        CoCoNut~\cite{lutellier2020coconut} & No        & Yes     & No       & No        \\
        CURE~\cite{jiang2021cure}           & No        & No      & -        & No        \\

        \bottomrule
    \end{tabular}
\end{table}

\subsection{How to modify them?}
\label{sec:RQ3_how_to_modify}

Due to the differences in design goals between APR and mutation tools,
we are forced to make a few modifications to the chosen APR tools:

\subsubsection{Where to fix (i.e., mutate)}
APR tools employ FL techniques to locate the buggy statements. In contrast,
mutation tools usually have manual options for specifying the files to be
mutated. Thus, we make APR tools to target the locations to be mutated by
directly manipulating the FL results.

\subsubsection{When to terminate}
for mutation testing, we assume that PUT has 
no defects, i.e., it has a green test suite. However, APR tools assume that PUT 
has defects, so they terminate when the candidate patch passes all tests. As 
our mutation goal is to simulate all possible mutants, we modify the 
termination criterion of APR tools so that they generate all target mutants 
without considering test results.

\subsubsection{Filtering some pre-defined patterns}
TBar is
based on the templates that have been collected from the fix patterns in the
repositories, as well as patches generated by other APR tools. Among the 
collected patterns, there are some patterns that are highly likely 
fix-patterns, e.g., inserting a null pointer checker. Even if it is possible that 
the developers would insert a wrong or inappropriate null pointer checker, we 
exclude such patterns by default, as we posit that those cases are rare. The 
two APR tools, TBar and PraPR, are modified in this way. However, to evaluate the 
effect of this filtering strategy, we also include the versions without such 
filtering, which we denote TBar$_{\alpha}$ and PraPR$_{\alpha}$.

\subsubsection{Sampling}
APR tools are generally designed to focus their efforts into a single location 
that is believed to be the location of the fault (which is why their results 
are highly dependent on the FL results~\cite{liu2019you}). In contrast, 
mutation tools aim to evenly spread their efforts across the entire program, 
with ways to sample mutants to control their numbers. In this regard, we modify 
APR tools so that we randomly sample only five mutations per location.

\subsection{Coupling Study}
\label{sec:RQ3_coupling_study}

Coupling Effect Hypothesis (CEH) states that, if a test suite can detect and
kill simple mutants, it will also be able to detect larger and more complex
faults~\cite{offutt1992investigations}. This is why mutation testing is supposed
to work. As such, the effectiveness of a mutation testing tool can be precisely
measured if we can measure the degree of coupling with real faults.

We follow the same procedure of the coupling study adopted by Just et
al.~\cite{just2014mutants}. The mutants are said to be \emph{coupled} with real
faults, if they are killed only by the test cases that reveal the real faults.
Given that there are failing test cases, $ft_i \in FT$, let $T_{pass}$ be a set
of passing tests and $T_{fail}$ be a set of tests that includes all tests in
$T_{pass}$ and a single failing test case, i.e., $T_{fail} = T_{pass} \cup
\{ft_i\}$. We then compose the pairs of $T_{pass}$ and $T_{fail}$ denoted by
$\braket{T_{pass}, T_{fail}}$ using the bugs and the corresponding failing test
cases in \d4j. For each pair, if there is at least one mutant that survives
$T_{pass}$ but is killed by $T_{fail}$, we mark the pair as \textit{coupled}. Generating
mutants using the modified APR tools, we compare the number of \textit{coupled pairs}
with the results from the two baseline mutation tools, PIT and Major.

Although we limit the mutant generation using random sampling at runtime, the
number of generated mutants varies significantly between the studied tools. It
would not be a fair comparison if there is a tool that generates many more
mutants than others, as it will by definition have a higher chance of being
coupled with real faults. Therefore, for any tool that generates more mutants
than our reference mutation tool, Major, we further take a random sample out of
those tools so that we consider the same number of mutants as Major. We repeat
the experiment 30 times and report mean of the numbers. We will also report the
unsampled total couplings, to see the impact of this additional sampling.

\subsection{Configurations}
\label{sec:RQ3_config}

We use the default set of mutation operators provided by Major in \d4j, and the
mutators in `old defaults group' for
PIT.\footnote{\url{https://pitest.org/quickstart/mutators/}} To compose
$\braket{T_{pass}, T_{fail}}$ pairs from bug benchmark, we use \d4j v1.2.0 and
exclude some subjects that we fail to run all subject tools on, resulting in 493
pairs. We ignore any equivalent mutants, as we are only interested in coupled
mutants, while equivalent mutants cannot be coupled by definition. For all APR
tools, we follow the settings specified in their original study. To alleviate a
huge cost of running tests against mutants, we generate mutants on
the files that the target faults reside in.

\begin{table}[ht!]
    \centering
    \caption{Results of a fixed-size fault coupling study}
    \label{tab:RQ3_results_sampled}
    \begin{tabular}{lrrr}
        \toprule
        Tool & \# Coupled Pairs & \# Total Pairs & \# Total Mutants  \\
        \midrule

Major & 300 & 493 & 31,877 \\
PIT & 249 & 493 & 29,855 \\
TBar & 316 & 493 & 31,679 \\
\inv{TBar} & 314 & 493 & 31,679 \\
TBar$_{\alpha}$ & 306 & 493 & 31,731 \\
SequenceR  & 90 & 493 & 29,291 \\
\inv{SequenceR} & 132 & 493 & 24,720 \\
SimFix & 99 & 493 & 29,411 \\
PraPR & 204 & 493 & 31,874 \\
PraPR$_{\alpha}$ & 175 & 493 & 31,877 \\
    
        \bottomrule
    \end{tabular}
\end{table}

\begin{figure}[!ht]
    \centering        
        \includegraphics[width=\linewidth]{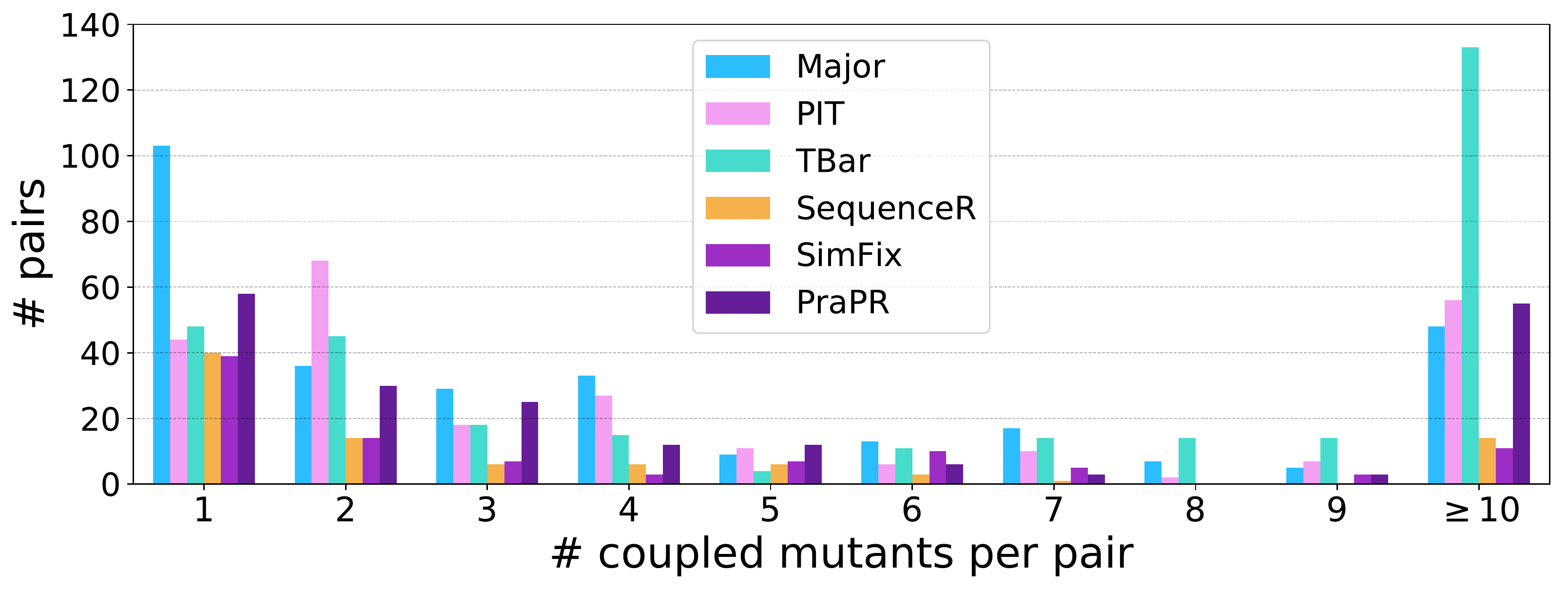}
    \caption{A distribution of the number of coupled mutants for each coupled pair.}
    \label{fig:RQ1_barchart}    
\end{figure}

\subsection{Results}
\label{sec:RQ3_results}


We begin by presenting the results of the fixed-size fault coupling study, for 
which we sample the same number of mutants as Major from all studied mutation 
tools. Subsequently, we present the results of unrestricted coupling study, for 
which we use all mutants generated by each tool. Finally, we present a 
qualitative analysis of the new types of mutations introduced by APR 
tools.

\subsubsection{Fixed-Size Coupling Study Results}

Table~\ref{tab:RQ3_results_sampled} presents the results of the fixed-size
coupling study: it shows the total number of pairs, the number of coupled pairs,
and the total number of mutants considered for each tool. TBar and its variants
perform better than two traditional mutation tools by making up to 316 coupled
pairs out of total 493 pairs (64\%). In addition, we examine the number of
coupled \emph{mutants} for each coupled pair as shown in
Figure~\ref{fig:RQ1_barchart}. On the pairs that have more than ten coupled
\emph{mutants}, TBar makes 133 coupled paired whereas Major and PIT make 48 and
56 coupled pairs.\footnote{Note that the duplicated
mutants~\cite{kintis2017detecting}, i.e., mutants that are semantically
different from the original, but equivalent to each other, may inflate these
numbers. However, since mutant equivalence is in general undecidable, we simply
report the raw results. We expect TBar to be largely free from this effect, as
each mutant it generates correspond to a unique template. We have excluded
syntactic duplicates.} It is well known that the fix templates of TBar have been
a powerful baseline for APR, and its effectiveness is shown to be valid for
mutant generation.


APR tools other than TBar, on the other hand, are outperformed by the two
mutation tools, Major and PIT. A closer inspection of the \emph{mutants}
generated by SequenceR and SimFix suggests that the degree of freedom these
tools have sometimes results in mostly frivolous code changes. Mutations created
by these tools include insertion of duplicate logical clauses (e.g., changing
\mintinline[escapeinside=||]{diff}{|\textcolor{codegrey}{\textbf{if}}| (a != 0)}
to \mintinline[escapeinside=||]{diff}{|\textcolor{codegrey}{\textbf{if}}| ((a !=
0) && (a != 0))}) or insertion of pointless parentheses (e.g., changing
\mintinline[escapeinside=||,
breaklines]{diff}{|\textcolor{codegrey}{\textbf{return}}| m !=
|\textcolor{codegrey}{\textbf{null}}| && m.containsKey(value)} to
\mintinline[escapeinside=||,
breaklines]{diff}{|\textcolor{codegrey}{\textbf{return}}| (m !=
|\textcolor{codegrey}{\textbf{null}}|) && (m.containsKey(value))}). We
hypothesise that these kinds of changes lack purpose, both as a patch and as a
mutation: compared to the carefully curated change templates of TBar, the
changes made by SequenceR are much more difficult to interpret as it depends on a
sequence-to-sequence NMT model. Since the NMT model does not concern semantics
of the produced token squences, some of the \emph{translations} may lack
purposes. Similarly, SimFix depends mostly on structural similarity to source
the change to be applied as mutation. Without considering the surrounding
semantic context, the changes are also likely to lack focus. 

Interestingly, PraPR performs worse than PIT, producing 204 coupled pairs 
compared to 249 made by PIT, although it is built based on PIT by augmenting 
nine mutation operators of PIT with six additional operators. However, it is 
the six newly introduced operators that mainly contribute to the performance 
deterioration. These operators perform either addition of field and method 
guards, or addition of pre/post conditions, which we think are too specialised 
for the purpose of program repair to perform as generic fault injection.



Finally, we investigate whether filtering pre-defined patterns has any merits
by comparing TBar with TBar$_{\alpha}$, and PraPR with PraPR$_{\alpha}$, 
respectively. Both comparisons show that the filtering helps generation of 
coupled mutants: TBar makes ten more coupled pairs, while PraPR makes 29 more. 

\begin{table}[ht]
    \centering
    \caption{Results of an unrestricted fault coupling study}
    \label{tab:RQ3_results_not_sampled}
    \begin{tabular}{lrrr}
        \toprule
        Tool & \# Coupled Pairs & \# Total Pairs & \# Total Mutants  \\
        \midrule

        Major & 300 & 493 & 31,877 \\
        PIT & 306 & 493 & 48,133 \\
        TBar & 416 & 493 & 83,185 \\
        \inv{TBar} & 413 & 493 & 82,943 \\
        TBar$_{\alpha}$ & 414 & 493 & 103,603 \\
        SequenceR  & 154 & 493 & 50,072 \\
        \inv{SequenceR} & 164 & 493 & 30,773 \\
        SimFix & 166 & 493 & 64,887 \\
        PraPR & 326 & 493 & 186,640 \\
        PraPR$_{\alpha}$ & 328 & 493 & 242,258 \\
    
        \bottomrule
    \end{tabular}
\end{table}

\subsubsection{Unrestricted Coupling Study Results}


Table~\ref{tab:RQ3_results_not_sampled} shows the results of a coupling study
using all generated mutants. TBar and its variants again outperform others by
making up to 416 coupled pairs out of 493 total pairs (84\%), showing that
pre-defined fix patterns are capable of generating effective mutants. SequenceR
and SimFix also perform better when we do not restrict the number of mutants:
they generate 64 and 67 more coupled pairs compared to the fixed-size coupling
study results. However, they are still significantly less effective than PIT and
Major.

While unrestricted PraPR and PraPR$_{\alpha}$ outperform PIT and Major, this is 
mainly thanks to the huge number of mutants they generate: PraPR$_{\alpha}$ 
generates 200k mutants, compared to 40k generated by PIT. The operators newly 
introduced to PraPR are responsible for these extra 160k mutants, but they only 
contribute 20 additional coupled pairs.

In the unrestricted coupling study, TBar$_{\alpha}$ and PraPR$_{\alpha}$ subsume
TBar and PraPR, respectively. Therefore, the effect of filtering some of the
pre-defined patterns can be clearly observed by comparing them: both
TBar$_{\alpha}$ and PraPR$_{\alpha}$ only make two more coupled pairs than TBar
and PraPR, respectively, despite generating 1.2 times more mutants.
Consequently, we conclude that the advantages of using all pre-defined patterns
are negligible.


\subsubsection{New and Stronger Mutation Operators}

\begin{figure}[!ht]
    \begin{subfigure}{\linewidth}
        \centering
\begin{minted}{diff}
+ for (int i = 0; i < listSize; i++) {            
+     if (!ShapeUtilities.equal ...
\end{minted}
        \caption{Chart-6 (fix)}
        \label{fig:chart_6_diff}
        \vspace{1em}
    \end{subfigure}

    \begin{subfigure}{\linewidth}
        \centering
\begin{minted}{diff}
- ... Partial(|\textcolor{codered}{\textbf{iChronology, newTypes, newValues}}|);            
+ ... Partial(|\textcolor{codegreen}{\textbf{newTypes, newValues, iChronology}}|);
\end{minted}
        \caption{Time-4 (fix)}
        \label{fig:time_4_diff}
        \vspace{1em}
    \end{subfigure}

    \begin{subfigure}{\linewidth}
        \centering
\begin{minted}{diff}
- ... convertLocalToUTC(localInstant, false);            
+ ... convertLocalToUTC(localInstant, false, |\textcolor{codegreen}{\textbf{instant}}|);
\end{minted}
        \caption{Time-26 (fix)}
        \label{fig:time_26_diff}
        \vspace{1em}
    \end{subfigure}
    
    \begin{subfigure}{\linewidth}
        \centering
\begin{minted}{diff}
- return |\textcolor{codered}{\textbf{allResultsMatch}}|(n, MAY_BE_STRING_PREDICATE);            
+ return |\textcolor{codegreen}{\textbf{anyResultsMatch}}|(n, MAY_BE_STRING_PREDICATE);
\end{minted}
        \caption{Closure-10 (fix)}
        \label{fig:closure_10_diff}
        \vspace{1em}
    \end{subfigure}
    
    \begin{subfigure}{\linewidth}
        \centering
\begin{minted}{diff}
+ removeDuplicateDeclarations(root);            
...
- removeDuplicateDeclarations(root);
\end{minted}
        \caption{Closure-102 (fix)}
        \label{fig:closure_102_diff}        
    \end{subfigure}

    \caption{The real faults in \d4j that are exclusively coupled with the mutants of APR tools.}
    \label{fig:qualitative_study}
\end{figure}

Just et al.~\cite{just2014mutants} reported that 27\% of real faults are not
coupled to the mutants generated by Major, and categorised the types of those 
uncoupled real faults. Based on it, we investigate the real faults to which PIT 
and Major cannot couple any mutants, but APR tools can, and report the 
operators used by the APR tools.

\begin{itemize}
\item \textit{Statement or code deletion}: TBar can delete a single or multiple 
lines of code (see Figure~\ref{fig:chart_6_diff}), whereas both PIT and Major 
lack the Statement Deletion (SDL) mutation operator to avoid compilation 
failures. 

\item \textit{Argument swapping}: TBar and PraPR can generate mutants that swap 
the arguments to the method call (see Figure~\ref{fig:time_4_diff}), as they 
anticipate swapped arguments as a potential developer mistake. 

\item \textit{Argument omission}: TBar, PraPR, and SequenceR can remove
extra arguments from a method call (see Figure~\ref{fig:time_26_diff}), anticipating such omissions as a potential mistake. 
    
\item \textit{Similar method called}: PraPR can change the method called to 
a similar method since PraPR has a method replacement operator that is 
not a part of PIT operators (see Figure~\ref{fig:closure_10_diff}), in anticipation of developer mistakes.
    
\item \textit{Statement shifting}\footnote{This type was not originally
listed by Just et al.~\cite{just2014mutants}.}: TBar can change the location
of a statement to the other line thanks to its move statement operator (see
Figure~\ref{fig:closure_102_diff}).
\end{itemize}




\begin{tcolorbox}[boxrule=0pt,frame hidden,sharp corners,enhanced,borderline north={1pt}{0pt}{black},borderline south={1pt}{0pt}{black},boxsep=2pt,left=2pt,right=2pt,top=2.5pt,bottom=2pt]
\textbf{Answer to RQ3:} APR tools can successfully generate the mutants coupled 
with real faults, revealing new and stronger mutation operators.
\end{tcolorbox}

\section{Threats to Validity}
\label{sec:threats}
Threats to internal validity concern any factor that may influence the observed 
effects. To mitigate such threats, we limit the APR and mutation tools we study 
to those that are publicly available and widely studied. We also limit any 
modification we introduce to the minimum. 

Threats to external validity concern the degree to which our results can be
generalised. We tried to incorporate as many datasets and programming languages
as possible, by studying C and JavaScript (RQ1) as well as Java (RQ2 and 3).
While our idea is not inherently confined to a specific programming language,
only further experimentations can generalise our results to new tools and
languages. We adopt \d4j as the standard benchmarks in both mutation testing and APR.

Threats to construct validity occur when the metrics we use fail to measure what
we initially plan to observe. For RQ1, we simply count the number of clusters as
there are no clear ground truths: we believe this is the simplest and the most
direct measurement we can take. For RQ2, we only report the number of bugs for
which subject tools can generate plausible patches. This is a simple count
based metric that can be validated by test execution. For RQ3, we report the
number of coupled pairs, following a widely accepted protocol.

\section{Related Work}
\label{sec:related_work}

Weimer et al.~\cite{Weimer2013ma} explored the duality between APR and mutation
testing. Specifically, they formalised and characterised Generate \& Validate
(G\&V) program repair as a dual of mutation testing: the equivalent mutant
problem is related to the redundant repairs, while the coupling effect is
related to the hypothesis that simple operators can fix many complex faults.
Both assume the competent programmer hypothesis, meaning that APR also assumes 
that a relatively simple patch can repair the bug. This is what PraPR explicitly
exploits using the mutation operators implemented in
PIT~\cite{ghanbari2019practical}. In this work, we further investigate the
approach taken by PraPR using \inv{TBar} and \inv{SequenceR}. Moreover, our work
is not confined to APR and mutation testing, but also considers the similarities
between human written patches and faults (RQ1). While Weimer et al. proposed the
duality as a theoretical framework, we present empirical evidence of the
relationship.

Brown et al.~\cite{brown2017care} proposed a mutant mining technique that 
essentially inverts fix changes mined from open source repositories. In 
one of their experiments, they ask whether `forward' and `backward' patches are 
different. The forward patches refer to the original fix changes, whereas 
backward patches refer to their inversions (i.e., faults). Interestingly, their 
results showed an overlap of 1,710 mined operators, out of 13,929 operators mined from both directions. However, all operators were mined 
from a single project, Space~\cite{wong1997test}, limiting the scope of generalisation. We have conducted 
a larger empirical evaluation with multiple programs and languages. Our results also suggest that mining mutation operators only from 
one direction may miss some relevant code changes.

While APR techniques can successfully patch many faults, it is known that they 
also produce many incorrect changes during the process~\cite{le2011genprog}. 
This partly motivates our use of APR tools as a source of code mutation. 
Recently proposed NMT based APR techniques seek to avoid the generation of 
incorrect and wasted patches by incorporating the surrounding contexts 
better~\cite{lutellier2020coconut, jiang2021cure}. The qualitative analysis of 
our results for RQ2 and 3 hints at the importance of contexts, calling for future work on 
ways to representing as well as comparing them.

\section{Conclusion \& Future Work}
\label{sec:conclusion}

This paper aims to relate two seemingly opposite concepts in software testing,
fixing (patch) and introducing bugs (fault). We highlight their syntactic
similarities based on empirical evaluations. An analysis of 13k fix- and
bug-inducing changes collected from open source repositories shows that, when
abstracted and clustered together, it is difficult to distinguish patches from
faults: up to 70\% of the patches and faults are clustered together. Based on
these results, we also show that mutation tools can be successfully used as APR
tools, and vice versa. An evaluation using \d4j bugs shows that mutation tools
generate plausible patches for 42 and 17 bugs, only eight and two fewer than
original APR tools, TBar and SequenceR, respectively. Finally, we also show that
APR tools can successfully generate mutants that are coupled with real faults.
Our findings suggest that the scope of code changes traditionally used to mine
mutation operators, or to learn fix patterns and templates, may need to be
widened to incorporate additional code changes that are relevant.

Future work could include a developer study where participants are asked to guess
whether the given code change is a patch or a fault, to further understand the
impact of context as well as a process of speculation. In addition, another
interesting direction could be exploring the potential of using our findings to
improve software testing tools. One example would be exploiting both directions
of code changes to mine the real faults in the open source projects. Previous
work has resorted to inverting the fix commits to generate the faults. This is
due to the fact that identifying bug-inducing commits is challenging, as the
developers may not be aware of introducing a bug and it is likely to contain the
changes unrelated to the bug~\cite{wen2019exploring, An2021qb}. On the other
hand, identifying fix commits is relatively straightforward, as the developers
intend to fix the bug with a commit message such as `This handles the bug
related to issue \#10'. Our finding suggests that by reversing the code changes
we can simply double the size of the existing dataset of real faults. A next
step could be to conduct a study that utilises this larger dataset, such as
comparing the performances of the learning-based static bug finders trained on
the original dataset and the larger one, respectively.

\section*{Acknowledgment}

This work is supported by National Research Foundation of Korea
(NRF) Grant (NRF-2020R1A2C1013629), Institute for Information \& communications
Technology Promotion grant funded by the Korean government (MSIT)
(No.2021-0-01001), and Samsung Electronics (Grant No. IO201210-07969-01). 

\newpage

\bibliographystyle{IEEEtran}

\bibliography{ref}
\balance

\end{document}